\documentclass[12pt]{iopart}
\usepackage[dvips]{graphicx}
\usepackage{latexsym}
\usepackage{iopams}
\usepackage{amssymb}
\usepackage{verbatim,times,bbm}

\newtheorem{definition}{Definition}[section]

\newtheorem{prop}[definition]{Proposition}

\newcommand{\al}{\alpha}

\newcommand{\vp}{\varphi}
\newcommand{\C}{\mathbb{C}}
\newcommand{\U}{U_q(\mbox{sl}_2)}

\newcommand{\Un}{U_q(\mbox{sl}_{n+1})}
\newcommand{\eee}{\mbox{exp}}
\newcommand{\hs}{\hfill $\square$}

\begin{document}

\title{Quantum state transfer in a $q$-deformed chain}

\author{Sonia L'Innocente$^{1,2}$, Cosmo Lupo$^2$, Stefano
Mancini$^2$}

\address{$^1$ Dipartimento di Matematica ed Informatica, Universit\`a di Camerino,
62032 Camerino, Italy}

\address{$^2$ Dipartimento di Fisica, Universit\`a di Camerino, 62032 Camerino,
Italy}

\eads{\mailto{sonia.linnocente@unicam.it},
\mailto{cosmo.lupo@unicam.it}, \mailto{stefano.mancini@unicam.it}}

\begin{abstract}

We investigate the quantum state transfer in a chain of particles
satisfying $q$-deformed oscillators algebra. This general algebraic
setting includes the spin chain and the bosonic chain as limiting
cases. We study conditions for perfect state transfer depending on
the number of sites and excitations on the chain. They are
formulated by means of irreducible representations of a quantum
algebra realized through Jordan-Schwinger maps. Playing with
deformation parameters, we can study the effects of nonlinear
perturbations or interpolate between the spin and bosonic chain.

\end{abstract}


\section{Introduction\label{intro}}

Spatially distributed interacting quantum systems can provide means
to transfer quantum information from one place to another. This
possibility relies on quantum interference effects arising from the
evolution of the whole system. An example along this line is given
by a chain of spin-$\frac{1}{2}$ systems where perfect state
transfer from one to another end can be realized
\cite{Bose,Christandl}. Another example is given by a chain of
harmonic oscillators \cite{bosonic_1,bosonic_2}. These two examples
come, under the mathematical point of view, from the realizations of
two different algebras (the Lie algebra su$(2)$ and the
Heisenberg-Weyl algebra) corresponding to fermionic and bosonic
commutation relations. These latter can be seen as two limit cases
of more general commutation relations involving deformed algebras
parameterized by one continuous parameter \cite{Kury, Kulish}. Due
to the increasing interest on the topic of state transfer in a chain
of quantum systems (see e.g.\ \cite{contemp}), it would be
interesting to investigate the state transfer in a more general
algebraic setting. In perspective, that could pave the way to a
systematic study of the role of algebraic structures in the problem
of state transfer. Moreover, the deformed algebraic setting can be
used as a formal way of describing nonlinear interaction in the
quantum chain.

We start by considering a chain of $n+1$ sites described by a
nearest-neighbor Hamiltonian of the kind
\begin{equation}\label{H_bose}
H = \sum_{j=1}^{n} J_{j} \, \frac{{a_j}^\dag a_{j+1} +
{a_{j+1}}^\dag a_j}{2}
\end{equation}
where $J_{j}$ are the coupling constants. The ${a_j}^\dag$, $a_j$ are ladder operators
whose algebraic properties  determine the nature of
the quantum chain. Their canonical commutation and anticommutation
relations respectively define a bosonic and a fermionic quantum
chain. Moreover the fermionic chain can be mapped, via the
Jordan-Wigner map, to a chain of spin-$1/2$ \cite{Yung}.

Here we consider a quantum chain of $q$-deformed oscillators.
Several kinds of deformed oscillator algebras have been introduced
and studied in literature. Here we are mainly concerned with the
complex (associative unital) algebra, called the (\emph{symmetric})
$q$-\emph{oscillator algebra} and denoted by $\mathcal{A}_q$
\cite{Mac}. Each site of the quantum chain is endowed with a copy of
$\mathcal{A}_q$, with four generators $a^\dag$, $a$, $q^N$, $q^{-N}$
subject to the relations
\begin{equation}\label{qdef1}
a {a}^\dag - q {a}^\dag a = q^{-N} .
\end{equation}
\begin{equation}\label{qdef2}
q^{-N} q^{N} = q^{N}q^{-N} =1, \quad  q^{N} a^\dag=q a^\dag q^{N},
\quad q^N a= q^{-1} aq^{N}.
\end{equation}
From (\ref{qdef1}), (\ref{qdef2}) the following properties can be
easily derived:
\begin{equation}
{a}^\dag a =[N],\quad a {a}^\dag =[N+1],
\end{equation}
where the notation $[N]$ indicates the $q$-number $N$, defined as:
\begin{equation}\label{qnumero}
[N]:=\frac{q^N -q^{-N}}{q-q^{-1}} .
\end{equation}
It is suitable to recall that the algebra $ \mathcal{A}_q$ is a
$*$-algebra with involution such that $a^*=a^\dag$ and
$(q^N)^*=q^N$. A key role is played by the representation $T$ of
$\mathcal{A}_q$ on a Hilbert space $\mathcal{H}$ with an orthonormal
basis $\{|m\rangle : m \in \mathbb{N} \}$, defined as
\begin{equation}\fl\label{Fock}
T(a)|m\rangle =\sqrt{[m]}|m-1\rangle , \quad T(a^\dag)|m\rangle
=\sqrt{[m+1]}|m+1\rangle , \quad T(N)|m \rangle = m |m\rangle.
\end{equation}
If $ \mathcal{D}$ denotes the dense linear subspace of $\mathcal{H}$
spanned by the vectors $|m\rangle$, then the representation $T$
becomes the Fock representation of the $q$-oscillator algebra $A_q$,
that is, the $*$-representation of the $*$-algebra $\mathcal{A}_q$
on $\mathcal{D}$.


For our investigation, we need to introduce the algebra $
\mathcal{A}_q^{\mbox{ext}}$ obtained by adjoining formally elements
$q^{N/2}$ and $q^{-N/2}$ to $ \mathcal{A}_q$. Then, a chain of two
sites can be represented by the tensor product
$\mathcal{A}_q^{\mbox{ext} \, \otimes 2}$ of two $q$-oscillator algebras
$\mathcal{A}_q^{\mbox{ext}}$ whose generators are denoted by $a_1 \,
a^{\dag}_1, \, q^{N_1/2}, \, q^{-N_1/2}, \, a_2 \, a^{\dag}_2, \,
q^{N_2/2}, \, q^{-N_2/2}$. It is relevant to note that every element
of the set $a_1 \, a^{\dag}_1, \, q^{\pm N_1/2}$ commutes with any
element from $a_2 \, a^{\dag}_2, \, q^{\pm N_2/2}$. The great
difference with the classical case is that the $q$-oscillator
algebra (generated by the deformed relations) does not realize any
matrix algebra but realizes, by the deformed Jordan-Schwinger map, a
suitable quantum algebra which constitutes our mathematical
framework. As a consequence, we will see that the relations for
perfect state transfer can be formulated by its irreducible
representations.

The paper is organized as follows. In Section \ref{quantuma} we
present the quantum algebra $\Un$ for $n \geq 1$ by discussing some
crucial properties and emphasizing its (deformed) Jordan-Schwinger
realization in terms of $q$-oscillator algebras. In Section
\ref{repres}, the irreducible representations of $\Un$ are presented
by composing the Jordan-Schwinger map with the Fock representation
of $ \mathcal{A}_q$. This framework allows us to represent the
physical system of the chain with $n+1$ sites. Section \ref{rel} is
devoted to the study of state transfer through a chain of
$q$-deformed oscillators. For the case of a chain of spin-$1/2$,
fermions, or bosons, Hamiltonian function with nearest-neighbor
interaction as (\ref{H_bose}) allows perfect state transfer if the
coupling constants $J_j$ are suitably chosen. We consider the
efficacy, for the issue of quantum state transfer, of one of this
choices in the case of a chain of $q$-deformed oscillators.
Conclusions and possible physical applications are drawn in Section
\ref{econcludo}.

\section{The quantum algebra $U_q({\rm sl}_{n+1})$}\label{quantuma}

Before of analyzing the issue of state transfer through a chain of
$q$-deformed oscillators, we fix our mathematical setting.
\bigskip\\
Let $q$ be a complex number such that $q \neq 0$ and $q^2 \neq 1$.
We first consider the quantized universal enveloping algebra $\U$ of
the Lie algebra $\mbox{sl}_2$ of all traceless $2 \times 2$ matrices
with coefficients in the field of complex numbers $\C$. $\U$ can be
described as the associative algebra with the unity over $\C$ with
four generators $E, \, F, \, K, K^{-1}$ satisfying the defining
relations
\begin{equation}\label{gener}
KK^{-1} = K^{-1}K  \, = \, 1, \quad KEK^{-1} \, = \,  q^2 E, \quad
KFK^{-1} \, = \,  q^{-2} F,
\end{equation}
\begin{equation}\label{brack}
[E,F] \, = \,\frac{K-K^{-1}}{q-q^{-1}}.
\end{equation}
It can be shown by induction that the relations (\ref{gener}) and
(\ref{brack}) imply for every positive integers $s$ and $t$ the
formulas
\begin{equation}\label{brackpowerF}
[E,F^t] \, = \, [t]F^{t-1} \,
\frac{Kq^{1-t}-K^{-1}q^{t-1}}{q-q^{-1}},
\end{equation}
\begin{equation}\label{brackpowerE}
[E^s,F] \, = \, [s]E^{s-1} \,
\frac{Kq^{s-1}-K^{-1}q^{1-s}}{q-q^{-1}}.
\end{equation}
A key property of the algebra $\U$ is that it carries a Hopf algebra
structure. Indeed, we can remind that there exists a unique Hopf
algebra structure on $\U$ with comultiplication $\Delta$, counit
$\varepsilon$, antipode $S$
\begin{equation}\fl\label{cHopf}
\Delta(E)=E \otimes K + 1 \otimes E, \quad \Delta(F)=F \otimes 1 +
K^{-1} \otimes F, \quad \Delta(K)=K \otimes K,
\end{equation}
\begin{equation}\fl\label{aHopf}
S(K)=K^{-1}, \quad S(E)= - E K^{-1}, \quad S(F)=-KF, \quad
\varepsilon(K)=1, \quad \varepsilon(E)=\varepsilon(F)=0  .
\end{equation}
From now on, we refer to this algebra endowed with the Hopf algebra
structure as the quantum algebra $\U$.
\\The quantum algebra $\U$ could be supposed to be a quantum
analogue of the enveloping algebra $U(\mbox{sl}_2)$ of the Lie
algebra $\mbox{sl}_2$. In fact, $\U$ shares two main properties with
the classical one: it has no zero divisors (see e.g.\
\cite[Proposition 1.8]{Jant}) and it has a Poincaré-Birkhoff-Witt
type basis (see e.g.\ \cite[$\S$ 3.1]{KliSch}), that is, $\U$ as
$\C$-vector space is generated by the basis $\{ E^s K^l F^t \, | \,
s,t \in \mathbb{N}-\{0\} , \, l \in \mathbb{Z} \}$.

Unfortunately we can not straightforwardly recover $U(\mbox{sl}_2)$
from $\U$ by setting $q = 1$ (as it happens at the level of
representation theory) but by considering the limit of $q
\rightarrow 1$ of a slight reformulation of $\U$ at least for $q$
not a root of unity (see e.g.\ \cite[Section 3.1.3]{KliSch}). For
our goals, it is relevant to equip $\U$ with an involution $*: \U
\rightarrow \U$ which turns $\U$ into a Hopf $*$-algebra, usually
called the real form of $\U$ and denoted (slightly abusing the
notation) again by $\U$.\\ The realization of $\U$ in terms of the
$q$-oscillator algebra $\mathcal{A}_q^{\mbox{ext} \, \otimes 2}$ (with
generators $a_1 {a_1}^\dag$, $q^{\pm N_1/2}$, $a_2 \, {a_2}^\dag$,
$q^{\pm N_2/2}$) can be allowed by the (deformed) Jordan-Schwinger
map $\mbox{JS}_q : \U \rightarrow \mathcal{A}_q^{\mbox{ext} \,\otimes 2}$
defined (similarly to the classical case) as:
\begin{equation}\label{Jsq}
\mbox{JS}_q(E) =a^\dag_1 a_2, \quad \mbox{JS}_q(F) =a^\dag_2 a_1,
\quad \mbox{JS}_q(K) =q^{(N_1-N_2)/2}
\end{equation}
By composing the (unique) algebra homomorphism $\mbox{JS}_q$ with
the Fock representation of $\mathcal{A}_q^{\mbox{ext} \,\otimes 2}$,
irreducible representations of $\U$ can be obtained. These
representations give the right setting where the relations for the
state transfer in a chain with two sites can be formulated. The same
thing can be repeated when we consider a chain with $n+1$ sites.
Hence, we are going on introducing the related quantum algebra, that
is, the universal enveloping algebra $\Un$ of the Lie algebra
$\mbox{sl}_{n+1}$ of all traceless $n \times n$ matrices.
\bigskip\\
First, consider the Lie algebra $\mbox{sl}_{n+1}$ for $n \geq 1$ and
the root system $\Phi$ of $\mbox{sl}_2$ with a basis $\Pi$ formed by
$n$ roots $\Pi=\{\al_1, \ldots , \al_n \}$. According to the scalar
product $( \cdot , \cdot )$ on the vector space generated by $\Phi$,
we have that $(\al, \al)=2$ for every (short) root $\alpha$ of
$\Phi$.
\\
The quantized enveloping algebra of $\mbox{sl}_{n+1}$  is a
$\C$-algebra $\Un$ with $4n$ generators $E_{\al_j}, \, F_{\al_j}, \,
K_{\al_j} , \, K^{-1}_{\al_j}$ with $j= 1, \ldots , n$ and
relations:
$$
\begin{array}{lr}
K_{\al_j}E_{\al_l}K^{-1}_{\al_i} =  q^2 E_{\al_l} \quad \mbox{and}
\quad K_{\al_j}F_{\al_l}K^{-1}_{\al_j}= q^{-2}F_{\al_l} & (j=l)
\\
K_{\al_j}E_{\al_l}K^{-1}_{\al_j}=  q^{-1} E_{\al_l} \quad \mbox{and}
\quad K_{\al_j}F_{\al_l}K^{-1}_{\al_j}= q  F_{\al_l} & (|j-l|=1)
\\
K_{\al_j}E_{\al_l}K^{-1}_{\al_j}=  E_{\al_l} \quad \mbox{and} \quad
K_{\al_j}F_{\al_l}K^{-1}_{\al_j}=  F_{\al_l} & (|j-l| \geq 2)
\smallskip\\
K_{\al_j}K_{\al_l} =K_{\al_l} K_{\al_j} \quad \mbox{and} \quad
E_{\al_j}F_{\al_l} - F_{\al_l} E_{\al_j}= \delta_{jl}
\frac{K-K^{-1}}{q-q^{-1}} &
\\
E_{\al_j}E_{\al_l} =E_{\al_l} E_{\al_j} \quad \mbox{and} \quad
F_{\al_j}F_{\al_l} =F_{\al_l} F_{\al_j} & (|j-l| \geq 2)
\smallskip\\
E^2_{\al_j}E_{\al_l} - (q+q^{-1}) \,
E_{\al_j}E_{\al_l}E_{\al_j} + E_{\al_l}E^2_{\al_j} =0 & (|j-l|=1)  \\
F^2_{\al_j}F_{\al_l} - (q+q^{-1}) \,
F_{\al_j}F_{\al_l}F_{\al_j} + F_{\al_l}F^2_{\al_j} =0 & (|j-l|=1)  \\
\end{array}
$$
When $n=1$, we obviously obtain the relations (\ref{gener}),
(\ref{brack}) of the quantized universal enveloping algebra of
$\mbox{sl}_2$. Equally to the case of $\U$, a Hopf algebra structure
is carried by $\Un$ which is so treated as quantum algebra: to
define the comultiplication, the antipode and the counit it is
enough to apply the same relations (\ref{aHopf}), (\ref{cHopf})
(described for $\U$) to the generators $E_{\al_j}, \, F_{\al_j}, \,
K_{\al_j} , \, K^{-1}_{\al_j}$, with $j=1, \ldots, n$. Furthermore,
we can endow $\Un$ with an involution $*: \Un \rightarrow \Un$ which
turns $\Un$ in a Hopf $*$-algebra.
\\It is worth to note that when $n>1$, it is always possible to
consider a subalgebra of $\Un$ which is isomorphic to $\U$. More
precisely, $\forall i$ the tuple of generators $(E_{\al_j}, \,
F_{\al_j}, \, K_{\al_j} , \, K^{-1}_{\al_j})$ satisfies the same
relations (\ref{gener}), (\ref{brack}) of $\U$, so we have for each
$\al_j \in \Pi$ the homomorphism $\U \rightarrow \Un$ that takes $E$
to $E_{\al_j}$, $F$ to $F_{\al_j}$, $K$ to $K_{\al_j}$ and $K^{-1}$
to $K^{-1}_{\al_j}$. Furthermore, this homomorphism will turn out to
be isomorphism onto its image (in $\Un$).
\\As in the case $n=1$, we can relate $\Un$ with the
$q$-oscillator algebra $\mathcal{A}_q^{\mbox{ext}}$. We consider the
tensor product $\mathcal{A}_q^{\mbox{ext} \,\otimes n+1}$ of $n+1$ copies
of $\mathcal{A}_q^{\mbox{ext}}$ whose set of generators is $\{ a_1
\, a^{\dag}_1, \, q^{\pm N_1/2}, \ldots , a_{n+1} \, a^{\dag}_{n+1},
\, q^{\pm N_{n+1}/2} \}$. As the case of $n=1$, a possible
Jordan-Schwinger realization of $\Un$ is achieved by mapping
\begin{equation}\fl
\mbox{JS}_q(E_{\al_j}) =a^\dag_j a_{j+1}, \quad
\mbox{JS}_q(F_{\al_j}) =a^\dag_{j+1} a_j, \quad
\mbox{JS}_q(K_{\al_j}) =q^{(N_{j}-N_{j+1})/2}, \quad j=1, \ldots n.
\end{equation}

\section{The representation theory of $U_q({\rm sl}_{n+1})$}\label{repres}

When a physical realization of the quantum algebra is considered,
its representation theory plays a crucial role. The representations
of the quantum algebra $\U$, are classified into three categories
according to the value of $q$:
\begin{enumerate}
\item $q$ is generic, that is, $q$ can take any value except $q =
0, \, \pm1$ and a root of unity, \item q is a root of unity, \item
$q=0$ (this case is also known as the crystal base).
\end{enumerate}
It is known that for $q$ generic, all finite dimensional
representations of $\U$ are completely reducible and the irreducible
ones are classified in terms of highest weights. In particular, they
can be regarded as deformation of the representations of the
classical $U(\mbox{sl}_2)$. When $q$ is a root of unity, the
representations of $\U$ become strikingly different from the
classical case. They are not completely reducible and some finite
dimensional representations are not the highest weight ones. \\
As to $\Un$, its simple finite dimensional representations of $\Un$
are very similar to those of $\mbox{sl}_{n+1}$ as long as $q$ is not
a root of unity. For $n=1$, we have clearly all information about
the simple representations of $\U$ (or equivalently $\U$-modules):
for all positive integer $m$, there exist exactly two simple
representations of $\U$ of dimension $m+1$ which correspond to each
simple modules over $\mbox{sl}_2$. In general, when $n \neq 1$, the
quantum algebra $\Un$ has $2^{|\Pi|}$ simple representations
corresponding to each simple module for $\mbox{sl}_{n+1}$. These
$2^{|\Pi|}$ modules arise from the choice
of $\Pi$ signs.\\
There exist different ways to describe the representations of $\Un$,
but for our interest in chains with $n+1$ sites, we use an approach
carrying to irreducible finite dimensional representations of $\Un$
by composing the Jordan-Schwinger realization with the Fock
representation of the algebra $\mathcal{A}_q^{\mbox{ext} \,\otimes n+1}$
(see also \cite[$\S$ 5.3.4]{KliSch}). \\

First, assume $q$ is not a root of unity. The Fock representation of
the algebra $\mathcal{A}_q^{\mbox{ext} \,\otimes n+1}$ acting on the
Hilbert space $\mathcal{H}^{\otimes \, n+1}$ with orthonormal basis
$|m_1, \ldots , _{n+1} \rangle$, is determined by the formulas
(\ref{Fock}). \\By the composition $\vp:= T \circ \mbox{JS}_q$, an
infinite dimensional representation of the quantum algebra $\U$ can
be formulated by linear operators on the space $
\mathcal{D}^{\otimes \, n+1}$ \smallskip
$$
\vp \, : \quad  \Un \,\stackrel{JS_q}{\longrightarrow }  \,
\mathcal{A}_q^{\mbox{ext} \,\otimes n+1} \; \stackrel{T}{\longrightarrow}
\; \mathcal{L}(\mathcal{D}^{\otimes \, n+1}). $$
\smallskip Furthermore, the basis elements $|m_1,
\ldots , m_{n+1} \rangle$ of $\mathcal{H}^{\otimes \, n+1}$ are
represented as follows:
$$
|m_1, \ldots, m_{n+1} \rangle = \frac{T(a^\dag_1)^{m_1}}{[m_1]!}
 \, \frac{T(a^\dag_2)^{m_2}}{[m_2]!} \cdot \ldots \cdot
 \frac{T(a^\dag_{n+1})^{m_{n+1}}}{[m_{n+1}]!}|0, \ldots ,0\rangle.
$$
So, the generators $E_{\al_j}$ and $F_{\al_j}$ of $\U$ for $j=1,
\ldots , n+1$ are mapped by $\vp$ in this manner:
\medskip
\begin{eqnarray}\label{infin}
\fl\vp(E_{\al_j})\, |m_1, \ldots, m_{n+1} \rangle  & = &
T(a^\dag_j)T(a_{j+1}) \, \frac{T(a^\dag_2)^{m_2}}{[m_2]!} \cdot
\ldots \cdot
 \frac{T(a^\dag_{n+1})^{m_{n+1}}}{[m_{n+1}]!}|0, \ldots , 0\rangle \medskip\\
\fl & = &\sqrt{[m_j +1] [m_{j+1}]}\, | m_1, \ldots , m_j +1,
m_{j+1}-1, \ldots,  m_{n+1}\rangle ,
  \nonumber \bigskip\\
\fl \vp(F_{\al_j})\, |m_1, \ldots, m_{n+1} \rangle & = & \sqrt{[m_j]
[m_{j+1} +1]}\, | m_1, \ldots , m_j -1, m_{j+1}+1, \ldots,
m_{n+1}\rangle. \nonumber
\end{eqnarray}
\medskip
For any positive integer number $m$, the linear subspace $S^m$
spanned by the basis elements $|m_1, \ldots, m_{n+1} \rangle $ with
$m_1 + m_2 + \ldots + m_{n+1}=m$ is invariant under the
representation $\vp$. So, the invariant subspace $S^m$ of $
\mathcal{H}^{ \otimes \, n+1}$ is generated by the vectors
$$ x_{m_1, m_2, \ldots , m_{n+1}} \, := \, |m_1, \ldots, m_{n+1}
\rangle.
$$
If we consider the Bargmann-Fock realization of $ \mathcal{A}_q$
(that is, a realization of the Fock representation on the Hilbert
space of entire holomorphic functions), then $S^m$ represents the
$\C$-vector space of all homogenous polynomials of $n+1$ variables
$X_1, X_2, \ldots ,X_{n+1}$ and degree $m$. \\
The restriction of $T$ to the invariant subspace $S^m$ is equivalent
to the irreducible finite dimensional representations $\vp_{n, m}$
of $\Un$, $\vp_{n, m}: \Un \rightarrow \mbox{End}(S^m)$ according to
that $\vp= \oplus_{m \in \mathbb{N}-\{0\}} \vp_{n,m}$. By the action
of $\vp$ given in (\ref{infin}), the generators $E_{\al_j}, \,
F_{\al_j}, \, K_{\al_j}$ (with $j=1, \ldots , n$) of $\Un$ act by
$\vp_{n,m}$ as follows:
\begin{eqnarray}
\fl E_{\al_j} \, x_{m_1, \ldots , m_{n+1}} & = & \left\{
\begin{array}{ll}
    \sqrt{[m_j +1] [m_{j+1}]}\, x_{m_1, \ldots , m_j +1, m_{j+1}-1, \ldots,  m_{n+1}},
 & \hbox{if}\, m_{j+1} >0; \\
0, & \hbox{if}\, m_{j+1}=0. \\
\end{array}
\right.
\nonumber \medskip\\
\fl F_{\al_j} \, x_{m_1, \ldots , m_{n+1}} & = & \left\{
\begin{array}{ll}
    \sqrt{[m_j][m_{j+1} +1]} \,x_{m_1, \ldots , m_j -1, m_{j+1}+1 , \ldots,  m_{n+1}},
 & \hbox{if} \, m_{j} >0; \\
0, & \hbox{if} \, m_{j}=0 . \\
\end{array}
\right. \nonumber \medskip\\
\fl K_{\al_i} \, x_{m_1,m_2, \ldots , m_{n+1}} & = & q^{m_1 -
m_{i+1}} \, x_{m_1,m_2, \ldots , m_{n+1}}. \label{orthE}
\end{eqnarray}
Every $x_{m_1, m_2, \ldots , m_{n+1}}$ is a weight vector and spans
every nonzero weight space in $S^m$ (which therefore has dimension
1). In particular, all $E_{\al_i}$ annihilate $\bar{x}_{m,0,0,
\ldots , 0}$. Up to the scalar multiplication this is the only
vector with this property. Hence, $S^m$ is an irreducible
representation of $\Un$ (for every $n \geq 1$).

Actually, the construction of the representation space $S^m$ holds
even if $q$ is a root of unity, but in general the irreducibility of
$S^m$ is lost. For instance, for $n=1$, if the order $d$ of $q$ is
bigger that $m+1$, then $S^m$ is simple and the map $\vp_{1,m}$ acts
in the same way described above; if $d$ is smaller that $m+1$, then
no simple finite dimensional representation exists; if $d=m+1$ we
should discuss other conditions.

\section{Deformed chains and perfect state transfer}\label{rel}

We are now able to approach the study of state transfer in a chain
of $q$-deformed oscillators. We consider the following protocol. The
ends of the quantum chain, i.e.\ the $1$st and the $(n+1)$th site,
are assigned respectively to the sender and the receiver. The
remaining $n-1$ oscillators constitute the communication channel.
The quantum chain is initialized in the vacuum state $|0\rangle
|0\rangle^{\otimes n-1} |0\rangle$, defined by $T(a_j) |0\rangle =
0$. The transfer protocol begins when the sender prepares her
oscillator in a qu$D$it state $|\psi\rangle = \sum_{m=0}^{D-1} c_m
|m\rangle$ where, according to the Fock representation (\ref{Fock}),
$|m\rangle = {K_m}^{-1/2} T(a_1^\dag)^m |0\rangle$, with
\begin{equation}
K_m = [m][m-1] \dots [2] [1].
\end{equation}
Then the quantum chain evolves according to the
chain Hamiltonian (\ref{H_bose}). Notice that the Hamiltonian
(\ref{H_bose}) preserves the total number of excitations in the
$q$-deformed chain. We refer to the manifold of states of the chain
with $m$ excitations as the $m$th Fock layer. It follows that the
chain dynamics does not mix Fock layer of different degree. After a
transfer time $t$ the sender instantaneously decouples the $(n+1)$th
oscillator from the rest of the chain. At this point, the receiver
can apply a suitable phase gate $\mathcal{U}=\sum_{m=0}^{D-1}
e^{i\phi_m} |m\rangle\langle m|$ on her oscillator to maximize the
transfer fidelity \cite{Bose,bosonic_2}. This local transformation
at the receiver site is independent on the state encoded by the
sender and is only determined by the chain Hamiltonian, its length,
and the transfer time $t$. The reduced state of the oscillator at
the receiver site is hence denoted $\rho(t)$. To evaluate the
quality of the state transfer, we consider the transfer fidelity
$F(t)= \langle\psi|\rho(t)|\psi\rangle$, averaged over all possible
input states.

In the classical case of a chain of spin-$1/2$, necessary and
sufficient conditions for obtaining a perfect state transfer have
been determined, see e.g.\ \cite{Kay} for a review. In particular,
it is possible to reach a perfect transfer if the coupling constants
in the Hamiltonian (\ref{H_bose}) are modulated according to
\begin{equation}\label{spin_c}
J_j = \lambda \sqrt{ j (n+1-j)}.
\end{equation}
In this way, the chain evolution is formally equivalent to a
rotation about the $x$-axis of a 'big spin' expressing a collective
degree of freedom of the quantum chain \cite{Christandl}. The same
choice of the coupling constants allows perfect state transfer in a
bosonic chain \cite{bosonic_2}. In this case, in each Fock layer the
chain evolution is equivalent to a rotation of a collective spin
about the $x$-axis. The perfect state transfer can be seen as a
consequence of the algebraic identity
\begin{equation}\label{algrel}
e^{i \pi S_x} S_- e^{-i \pi S_x} = S_+,
\end{equation}
where $S_+$, $S_-$, $S_z$ are the collective spin operators in the
$m$th Fock layer, and the transfer time is independent of the length
of the chain and of the order of the Fock layer and equals
$t=\pi/\lambda$.


Using the theory of representations of $U_q({\rm sl}_{n+1})$ one can
explicitly show that the choice of the coupling constants
(\ref{spin_c}) allows perfect state transfer in a chain of
$q$-deformed oscillators if quantum information is encoded using
only the vacuum state and the first Fock layer. However, if higher
Fock layer are included in the encoding the choice (\ref{spin_c}) is
no longer sufficient to allow perfect state transfer in a chain of
$q$-deformed oscillators. Indeed, the effects of nonlinearity
introduced by the $q$-deformation manifest themselves if two or more
excitations are present in the quantum chain.

\subsection{PST in the first Fock layer}

Here we consider the case of the transfer of a qubit state encoded
as $|\psi\rangle = c_0 |0\rangle + c_1 |1\rangle$. In this
case, the chain dynamics only involves the vacuum state and the
first Fock layer.

Let us start to discuss the case when $n$, $m$ are both equal to 1,
that is, we have a network with two sites (so the quantum algebra
$\U$ as the mathematical model) and just one excitation. Thus, by
considering the representation map $\vp_{1, \, 1}: \U \rightarrow
S^1$, the matrices determined by the action (by $\vp_{1, \, 1}$) of
generators of $\U$
$$
\vp_{1,1}(E)= \left(
\begin{array}{cc}
  0 & 1 \\
  0 & 0 \\
\end{array}
\right), \, \vp_{1,1}(F)=\left(
\begin{array}{cc}
  0 & 0 \\
  1 & 0 \\
\end{array}
\right), \, \vp_{1,1}(K)= \left(
\begin{array}{cc}
  q & 0 \\
  0 & q^{-1} \\
\end{array}
\right)
$$
coincide with the generators of $\mbox{sl}_2$. As in the classical
case (see \cite{Christandl}), let us chose three variable $S_x$ and
$S_y$ in $\U$ as follows:
$$S_x := \frac{E+F}{2}, \quad S_y :=
\frac{E+F}{2i} \, ,$$
 and $S_{+}, S_{-} \in \U$ as:
$$
S_{+}:= S_x + i  S_y , \hspace{2cm} S_{-}:= S_x - i  S_y .
$$
By applying the representation map to these new variables, we can
easily note that $\vp_{1,1}(S_x)$, $\vp_{1,1}(S_y)$,
$\vp_{1,1}(S_z)$ coincide with the generators of the Lie algebra
$su(2)$ of traceless skew-hermitian matrices and $\vp_{1,1}(S_{+})$,
$\vp_{1,1}(S_{-})$ with the Pauli matrices, that is, with the
generators of the (special unitary) Lie group $SU(2)$ of unitary
matrices with unit determinant.


We now consider the case of a chain of $n +1$ $q$-deformed
oscillators.
The related mathematical setting is formed by the quantum algebra
$\Un$ with the generators $E_{\al_j}, \, F_{\al_j}, \, K_{\al_j}$
(for $j=1, \ldots , n$) and by the representation $S^m$ of all
homogenous polynomials of $n+1$ variables and degree $m$. A possible
strategy is that of generalizing the previous result shown for $n,
\, m=1$ to this framework. First, we can show the analogous
relations (\ref{algrel}) for the case of $n+1$ sites and 1
excitation (with $S^1$ the related representation).
\begin{prop}\label{rel1n}
Let $\vp_{n,1}$ denote the representation map $\vp_{n,1}: \U
\rightarrow \mbox{End}(S^1)$ taking the generators of $\Un$,
$E_{\al_j}, \, F_{\al_j}, \, K_{\al_j}$, respectively to the $(n+1)
\times (n+1)$ matrices $\vp_{n,1}(E_{\al_j}), \,
\vp_{n,1}(F_{\al_j}), \, \vp_{n,1}(K_{\al_j}) \in M_{n+1}(\C)$. \\
Let us set $S_x, S_y\in \Un$ as:
\begin{equation}\label{Sxyn}
\fl S_x:= \sum_{j=1}^n \sqrt{j(n-j+1)} \, \frac{E_{\al_j} +
F_{\al_j}}{2}, \qquad S_y:= \sum_{j=1}^n \sqrt{j(n-j+1)} \,
\frac{E_{\al_j} - F_{\al_j}}{2 i},
\end{equation}
and $S_{+}, S_{-} \in \Un$ as:
\begin{equation}\label{Spmn1}
S_{+}:= S_x + i  S_y , \hspace{2cm} S_{-}:= S_x - i  S_y
\end{equation}
Then, the relation
\begin{equation}
\eee(it\vp_{n,1}(S_x)) \, \vp_{n,1}(S_{-}) \, \eee(-it
\vp_{n,1}(S_x)) = \vp_{n,1}(S_{+})
\end{equation}
holds for the time value $t=\pi$.
\end{prop}
\indent \emph{Proof.} According to the relations (\ref{orthE})
applied to the $n+1$ basis vectors of $S^1$, $x_{1,0, \ldots, 0},
\ldots ,x_{0,0, \ldots, 1}$, the matrices $\vp_{n,1}(E_{\al_i}), \,
\vp_{n,1}(F_{\al_i}), \, \vp_{n,1}(K_{\al_i})$ are:
$$
 \vp_{n,1}(E_{\al_1})=\left(
\begin{array}{cccc}
  0 & 1 &  0 & \ldots \, 0\\
  0& 0 & \ldots &  0\\
  \vdots & \vdots & \ddots &  \\\
 0&  & \ldots &  0\\
  0&  & \ldots &  0\\
\end{array}
 \right),
  \, \ldots \, ,
\vp_{n,1}(E_{\al_n})=\left(
\begin{array}{cccc}
  0 & 0 &  \ldots &  0\\
  0& 0 & \ldots &   0\\
  \vdots & \vdots & \ddots &  \\\
 0& 0 & \ldots &  1\\
  0&  & \ldots &  0\\
\end{array}
 \right)
$$
$$
  \vp_{n,1}(F_{\al_1})= \left(
\begin{array}{cccc}
  0 & 0 &  \ldots &  0\\
  1& 0 & \ldots &  0\\
   0 &  & \ldots &  0\\
    \vdots &  &  & \vdots \\\
    0&  & \ldots &  0\\
\end{array}
\right), \ldots , \vp_{n,1}(F_{\al_n})= \left(
\begin{array}{cccc}
  0 & 0 &  \ldots &  0\\
  0& 0 & \ldots &  0\\
   0 & 0 & 0 \ldots &  0\\
    \vdots &  &  & \vdots \\
    0&  \ldots & 1 &  0\\
\end{array}
\right),
$$
$$ \vp_{n,1}(K_{\al_1})=\left(
\begin{array}{cccc}
  q & & \ldots & 0 \\
  0& q^{-1} & \ldots & 0 \\
  \vdots &  &  & \vdots \bigskip\\
   0& 0 & 1& 0 \\
    0 & 0 & \ldots & 1 \\
\end{array}
\right) , \ldots , \vp_{n,1}(K_{\al_n})=\left(
\begin{array}{cccc}
  1 & & \ldots & 0 \\
  0& 1 & \ldots & 0 \\
  \vdots &  &  & \vdots \bigskip\\
    0& 0 &  \ldots q & 0 \\
0 & 0 & \ldots & q^{-1} \\
\end{array}
\right).
$$
By choosing $S_x$, $S_y$ as in (\ref{Sxyn}) and $S_{+}$, $S_{-}$ as
in (\ref{Spmn1}), the corresponding matrices
\begin{eqnarray}
\vp_{n,1}(S_+)= \left(
\begin{array}{ccccc}
0      & \sqrt{n} & 0                             & \ldots & 0\\
0      & 0        & \hspace{-0.4cm} \sqrt{2(n-1)} & \ddots & 0\\
\vdots & \vdots   & \ddots                        & \ddots & \vdots\\
0      & 0        & 0                             & \ddots & \sqrt{n}\\
0      & 0        & 0                             & \ldots      & 0 \\
\end{array} \right), \, \label{Spmm_1} \\
\vp_{n,1}(S_-)=\left(
\begin{array}{ccccc}
0        & 0             & 0        & \ldots   & 0\\
\sqrt{n} & 0             & 0        & \ldots   & 0\\
0        & \sqrt{2(n-1)} & \ddots   & \ddots   & 0 \\
\vdots   & \ddots        & \ddots   & \ddots   & \vdots\\
0        & 0             & \ldots   & \sqrt{n} & 0
\end{array}
 \right) \label{Spmm_2}
\end{eqnarray}
are shown to be compatible with the classical case, so the statement
is easily proved. \hs

\subsection{State transfer in higher Fock layer}

Here we consider the case of a qu$D$it encoding exploiting states
which higher number of excitations. We study the transfer of one
qutrit encoded at the sender site in a state of the form
$|\psi\rangle = c_0 |0\rangle + c_1 |1\rangle + c_2 |2\rangle$ and
numerically evaluate the average transmission fidelity as function
of the transfer time and the deformation parameter, when the
coupling constants are chosen according to (\ref{spin_c}). For $q=1$
the 'classical' bosonic chain is recovered, and the choice of
coupling constants is optimal. Deviations from this classical
behavior appear as long as $q \neq 1$. The $q$-deformation in the
algebraic structures induces a nonlinear perturbation in the
spectrum of the bosonic chain. The nonlinear effects manifest
themselves when two or more excitations are present in the quantum
chain. This will in general affect the fidelity of the state
transfer with respect to the undeformed bosonic chain.

\begin{figure}[htbp]
\centering
\includegraphics[width=0.5\textwidth]{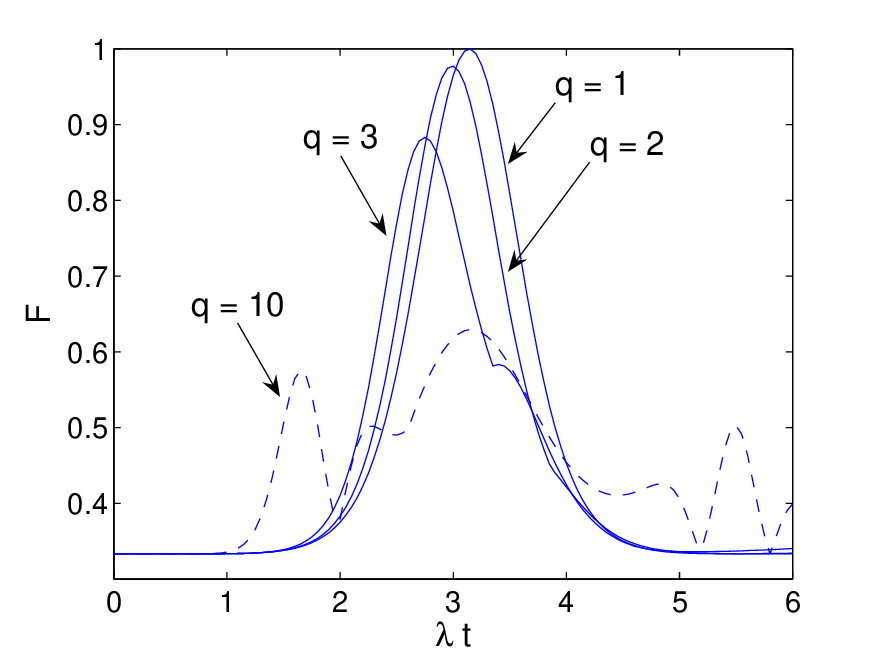}
\caption{The plot shows the average fidelity of the state transfer
versus the strength of the interaction $\lambda t$ in the second
Fock layer, for a chain of $10$ $q$-deformed bosons. Different lines
refer to different values of the deformation parameter. Notice that
the dynamics is symmetric under the exchange $q \leftrightarrow
q^{-1}$.} \label{t_nn_root}
\end{figure}

\begin{figure}[htbp]
\centering
\includegraphics[width=0.5\textwidth]{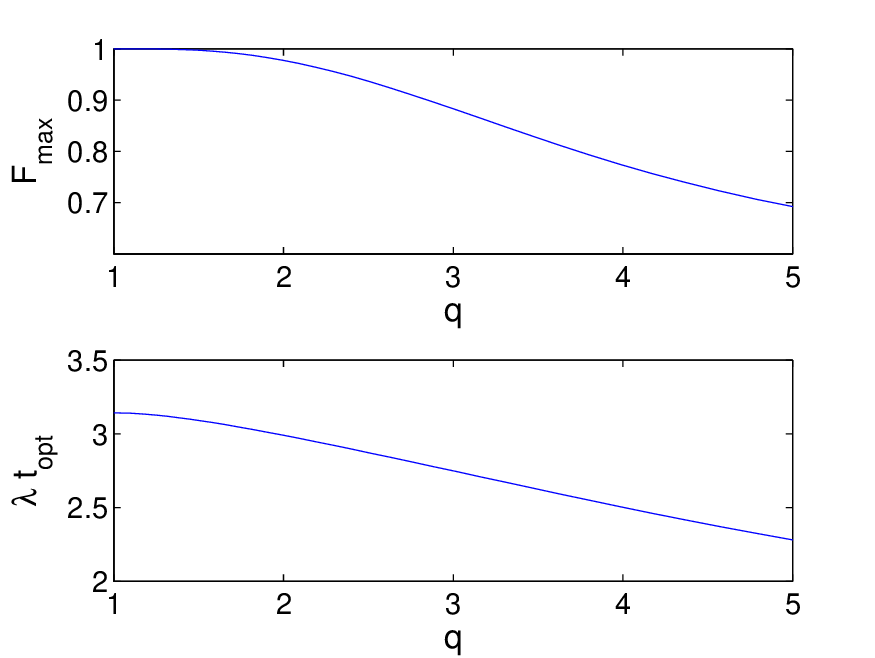}
\caption{For a chain of $10$ $q$-deformed bosons, the figure shows
the maximum average fidelity (top) of the state transfer in the
second Fock layer and the corresponding optimal (adimensional)
transfer time $\lambda t^*$ (bottom), as function of the deformation
parameter $q$. The analysis is restricted to a temporal window
$\lambda t \in [0,2\pi]$, corresponding to the period of the
undeformed dynamics \cite{bosonic_2}.} \label{q_nnroot}
\end{figure}

Figures \ref{t_nn_root} shows the average transfer fidelity as
function of the (adimensional) transfer time $\lambda t$, for a
chain of $10$ $q$-deformed oscillators. The undeformed chain,
recovered for $q=1$, allows perfect state transfer after a minimal
transfer time $\lambda t = \pi$. For increasing value of the
nonlinearity parameter $q$, the maximum average fidelity decreases,
while the (non-perfect) state transfer is generally faster. Figure
\ref{q_nnroot} shows the maximum average fidelity of the state
transfer and the corresponding optimal transfer time as function of
the deformation parameter. The analysis is restricted to a temporal
window $\lambda t \in [0,2\pi]$, corresponding to the period of the
undeformed dynamics \cite{bosonic_2}. Notice that, from the form of
the $q$-number (\ref{qnumero}), the dynamics is symmetric under the
exchange $q \leftrightarrow q^{-1}$.


\begin{figure}[htbp]
\centering
\includegraphics[width=0.5\textwidth]{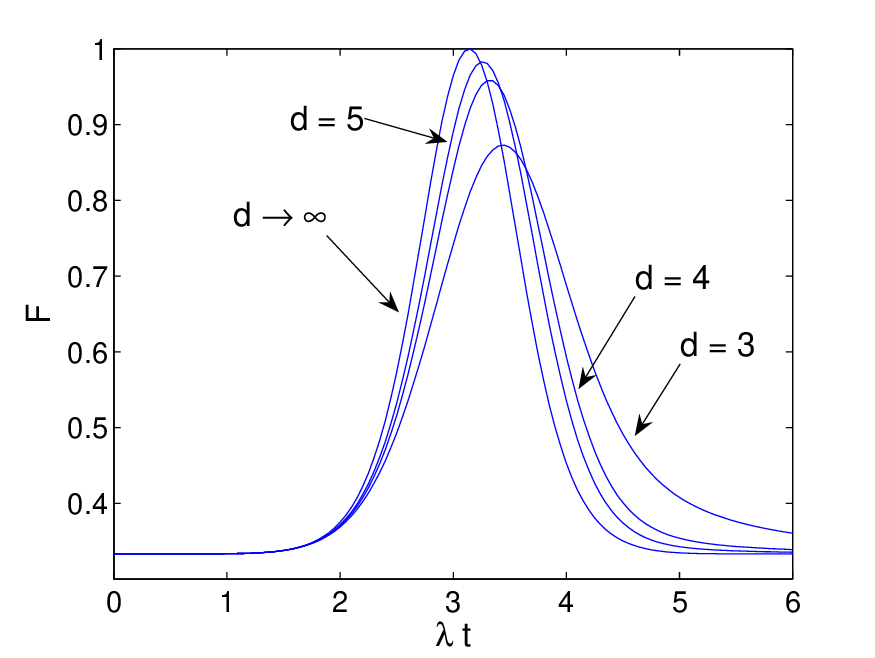}
\caption{The plot shows the average fidelity of the state transfer
versus the strength of the interaction $\lambda t$ in the second
Fock layer, for a chain of $10$ $q$-deformed bosons. The deformation
parameter is $q=e^{i\pi/d}$. Different lines refer to different
values of the deformation parameter. Notice that the classical
bosonic case is recovered in the limit $d\rightarrow\infty$.}
\label{t_root}
\end{figure}

\begin{figure}[htbp]
\centering
\includegraphics[width=0.5\textwidth]{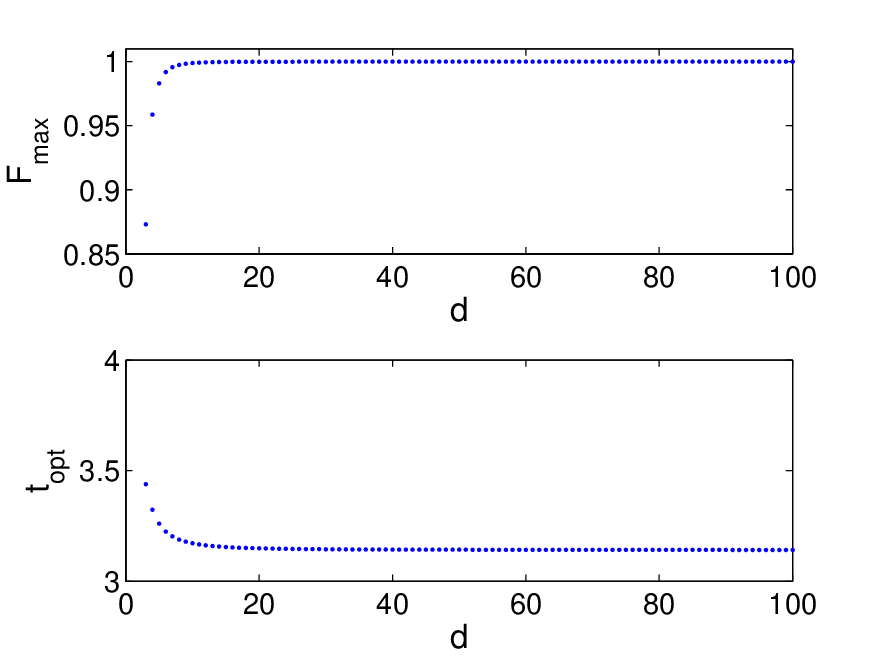}
\caption{For a chain of $10$ $q$-deformed bosonic oscillators with
$q = e^{i \pi/d}$, the figure shows the maximum average fidelity
(top) in the second Fock layer, and the corresponding optimal
(adimensional) transfer time $\lambda t^*$ (bottom), as function of
the deformation parameter $d$. Notice that the classical bosonic
case is recovered in the limit $d\rightarrow\infty$.} \label{d_root}
\end{figure}

In some cases the introduction of the $q$-deformation at the
algebraic level can be used to interpolate, varying the value of the
deformation parameter $q$, between the 'classical' cases of a chain
of spin-$1/2$ and a bosonic chain. For instance by choosing
$q=e^{\pm i \pi/d}$, for any integer $d$, it is possible to show
that the Fock space is the direct sum of $d$ dimensional subspace,
which are not connected by the ladder operators \cite{Chai}. This is
a consequence of the deformed commutation relations, which implies
$T(a_k)^d=0$, $T({a_k}^\dag)^d=0$. From this point of view, one can
consider the chain of deformed oscillators with $q = \exp{(\pm i
\pi/d)}$ as a chain of $d$-level systems with non-equally spaced
energy levels. Hence, by varying the integer $d$, one can
interpolate between the spin-$\frac{1}{2}$ case, obtained for $d=2$,
and the bosonic case, recovered in the limit of $d \rightarrow
\infty$. We consider the case $d > 2$, since for $d=2$ the condition
$T(a^\dag)^2=0$ (Pauli principle) avoids two excitations on the same
site. Figure \ref{t_root} shows the average fidelity of the state
transfer for a chain of $q$-deformed oscillators as a function of
the transfer time, for several value of the effective Hilbert space
dimension $d$. The minimal dimension in which the two-excitation
encoding can be defined is $d=3$. Notice that the bosonic limit is
recovered for $d\to\infty$, in which case perfect state transfer
happens for a minimal transfer time $\lambda t = \pi$. Finite values
of $d$ lead to a smaller transfer fidelity and a longer optimal time
transfer. Figure \ref{d_root} shows the maximum average transfer
fidelity and the corresponding optimal transfer time as function of
the effective Hilbert space dimension $d$.



\section{Conclusions}\label{econcludo}

We have considered the issue of state transfer through a quantum
chain of $q$-deformed oscillators. For real values of the
deformation parameter the physical consequence of the algebraic
deformation is the appearance of nonharmonicity in the energy
spectrum of the chain. The $q$-deformation can be hence interpreted
as a formal way to describe a bosonic chain with nonlinear
interactions. If only states with one excitation are involved the
nonlinearities do not play any role and the $q$-deformed dynamics is
identical to its classical, linear, counterpart. More generally, if
the considered protocol involves states of the chain with two or
more excitations, we have found that the nonlinear effects decrease
the fidelity of the state transfer, while however shortening the
optimal transfer time. Similar results were recently presented in
\cite{bosonic_2}, where the state transfer through a bosonic chain
described by the (nonlinear) Bose-Hubbard Hamiltonian was
considered. In our analysis we have chosen the coupling constants
according to (\ref{spin_c}), a choice which is optimal in the
undeformed case. Clearly, alternative $q$-dependent choices of the
coupling constants could lead to better performances.

Finally, if the deformation parameter is chosen to be a root of the
unity of order $d$ the $q$-deformed oscillator can be used to
simulate a $d$-level quantum system with nonequally spaced
stationary level. In this case, varying the deformation parameter
from $d=2$ to $d\to\infty$ one can describe a family of quantum
chain interpolating between a chain of spin-$1/2$ and the bosonic
chain.


\ack The work of C.L. and S.M. is partially supported by EU through
the FET-Open Project HIP (FP7-ICT-221899).

\end{document}